\documentclass[aps,pra,floatfix,superscriptaddress,twocolumn,showpacs,10pt]{revtex4-1}
\usepackage{amssymb, amsmath,color,mciteplus,graphicx,subfigure}
\usepackage{calc,epsfig,epstopdf,color,mciteplus,bm,mathrsfs}
\usepackage{times}
\usepackage{float}
\usepackage{multirow}
\usepackage{lipsum}

\begin{document}

\title{ Scalable protocol to mitigate $ZZ$ crosstalk in universal quantum gates}

\author{Yan Liang}
\affiliation{Key Laboratory of Atomic and Subatomic Structure and Quantum Control (Ministry of Education), Guangdong Basic Research Center of Excellence for Structure and Fundamental Interactions of Matter,\\ and School of Physics, South China Normal University, Guangzhou 510006, China}
\affiliation{School of Physical Science and Technology, Guangxi Normal University, Guilin 541004, China}

\author{Ming-Jie Liang}
\affiliation{Key Laboratory of Atomic and Subatomic Structure and Quantum Control (Ministry of Education), Guangdong Basic Research Center of Excellence for Structure and Fundamental Interactions of Matter,\\ and School of Physics, South China Normal University, Guangzhou 510006, China}

\author{Sai Li}
\affiliation{Key Laboratory of Atomic and Subatomic Structure and Quantum Control (Ministry of Education), Guangdong Basic Research Center of Excellence for Structure and Fundamental Interactions of Matter,\\ and School of Physics, South China Normal University, Guangzhou 510006, China}
\affiliation{Guangdong Provincial Key Laboratory of Quantum Engineering and Quantum Materials,\\   Guangdong-Hong Kong Joint Laboratory of Quantum Matter,  and Frontier Research Institute for Physics,\\ South China Normal University, Guangzhou 510006, China}

\author{Z. D. Wang} \email{zwang@hku.hk}
\affiliation{Guangdong-Hong Kong Joint Laboratory of Quantum Matter, Department of Physics,\\ and HK Institute of Quantum Science \& Technology,   The University of Hong Kong, Pokfulam Road, Hong Kong, China}

\author{Zheng-Yuan Xue}\email{zyxue83@163.com}
\affiliation{Key Laboratory of Atomic and Subatomic Structure and Quantum Control (Ministry of Education), Guangdong Basic Research Center of Excellence for Structure and Fundamental Interactions of Matter,\\ and School of Physics, South China Normal University, Guangzhou 510006, China}

\affiliation{Guangdong Provincial Key Laboratory of Quantum Engineering and Quantum Materials,\\    Guangdong-Hong Kong Joint Laboratory of Quantum Matter,  and Frontier Research Institute for Physics,\\ South China Normal University, Guangzhou 510006, China}
\affiliation{Hefei National Laboratory,  Hefei 230088, China}

\date{\today}

\begin{abstract}
High-fidelity universal quantum gates are widely acknowledged as essential for scalable quantum computation. However, in solid-state quantum systems, which hold promise as physical implementation platforms for quantum computation, the inevitable $ZZ$ crosstalk resulting from interqubit interactions significantly impairs quantum operation performance. Here we propose a scalable protocol to achieve $ZZ$-crosstalk mitigation in universal quantum gates. This method converts the noisy Hamiltonian with $ZZ$ crosstalk into a framework that efficiently suppresses all $ZZ$-crosstalk effects, leading to ideal target quantum operations. Specifically, we first analytically derive the $ZZ$-crosstalk mitigation conditions and then apply them to enhance the performance of target universal quantum gates. Moreover, numerical simulations validate the effectiveness of $ZZ$-crosstalk mitigation when multiple qubit gates operate concurrently.
As a result, our protocol presents a promising approach for implementing practical parallel quantum gates in large-scale quantum computation scenarios.

\end{abstract}
\maketitle

\section{Introduction}

Quantum computation is an emerging technology that leverages the principles of quantum mechanics to tackle problems that are intractable for classical computation, such as factorization of large integers \cite{Vandersypen2001,Lucero2012} and database searching \cite{Jones1998}. The success of quantum computation relies on the implementation of high-fidelity quantum operations. However, quantum crosstalk in large-scale quantum systems influences parallel quantum operations, leading to error accumulation and propagation \cite{Shahandeh2021}, which can ultimately cause the quantum computation process to fail. The $ZZ$ crosstalk resulting from interqubit interactions is prevalent in various quantum systems, such as semiconductor and superconducting qubits \cite{Buterakos2018, Throckmorton2022, AKandala2021, PZhao2022}, and leads to correlated and nonlocal errors \cite{LPostler2018,UvonLupke2020}, as well as spectator errors \cite{Sundaresan2020,Krinner2020,TQCai12021}. Therefore, the development of methods to implement universal quantum gates that can withstand
the effects of $ZZ$ crosstalk is crucial, particularly in quantum systems where that effect is significant.

Recently, significant efforts have been devoted to mitigating the $ZZ$-crosstalk effect in quantum information processing. Firstly, various hardware-based strategies have been proposed, including the integration of tunable couplers or buses \cite{PMundada2019,XYHan2020,LDuan2020,MCCollodo2020,JChu2021,pZhao2021,JStehlik2021,Youngkyu2021}, and the utilization of qubits with opposite anharmonicity \cite{JKu2020,PZhao2020,XXu2021}. However, these strategies heavily rely on the precision of hardware manufacturing, posing substantial challenges. Secondly, quantum control strategies, such as the simultaneous ac Stark effect on coupled qubits \cite{Noguchi2020,Mitchell2021,HXiong2022,KXWei2021,ZCNi2022} and dynamical decoupling \cite{Viola1999,Suter2016,Jurcevic2021,Tripathi2022}, offer alternative approaches that can alleviate hardware requirements while suppressing $ZZ$ crosstalk.  However, due to the lack of freedom of control, the extensibility of the Stark method \cite{Noguchi2020,Mitchell2021,HXiong2022,KXWei2021} is limited. Besides, the dynamical decoupling method needs to apply the decoupling pulse sequence independently to all spectators to eliminate the 
$ZZ$-crosstalk errors \cite{Tripathi2022}, resulting in more resource consumption. Recently, an analytical condition for achieving crosstalk-robust control has been proposed \cite{Zeyuan2022}, which is  applicable only to single-qubit cases.

In this paper, we address the above obstacles by presenting a scalable protocol for implementing universal quantum gates with $ZZ$-crosstalk mitigation (ZZCM) in large-scale quantum systems. The quantum system with a $ZZ$-crosstalk Hamiltonian is transformed into a framework that suppresses all $ZZ$-crosstalk effects between qubits, yielding ideal quantum operations. We analytically derive the conditions that the transformed operator must satisfy and apply it to enhance the performance of universal quantum gates. We demonstrate that, for single-qubit gates, the application of a continuous external drive field to the gate qubit can effectively suppress $ZZ$ crosstalk from all  nearby qubits, significantly reducing experimental complexity and yielding high-fidelity quantum gates. For parallel quantum gates, the mitigation of $ZZ$ crosstalk between adjacent qubits can be achieved by applying continuous external driving fields only to the next-nearest-neighbor qubits. Furthermore, our scheme does not need additional physical qubits or circuits and is suitable for all quantum processors. Consequently, this approach holds great promise for practical large-scale parallel quantum computation.

\section{Quantum Gates with ZZCM } \label{sec2}

In this section, we first present the considered lattice model for universal quantum gates with specification on the explicit form of the $ZZ$ crosstalk. Then, we propose the general scheme for constructing universal quantum gates with ZZCM.

\subsection{$ZZ$-Crosstalk Model}

\begin{figure}[tbp]
  \centering
  \includegraphics[width=\linewidth]{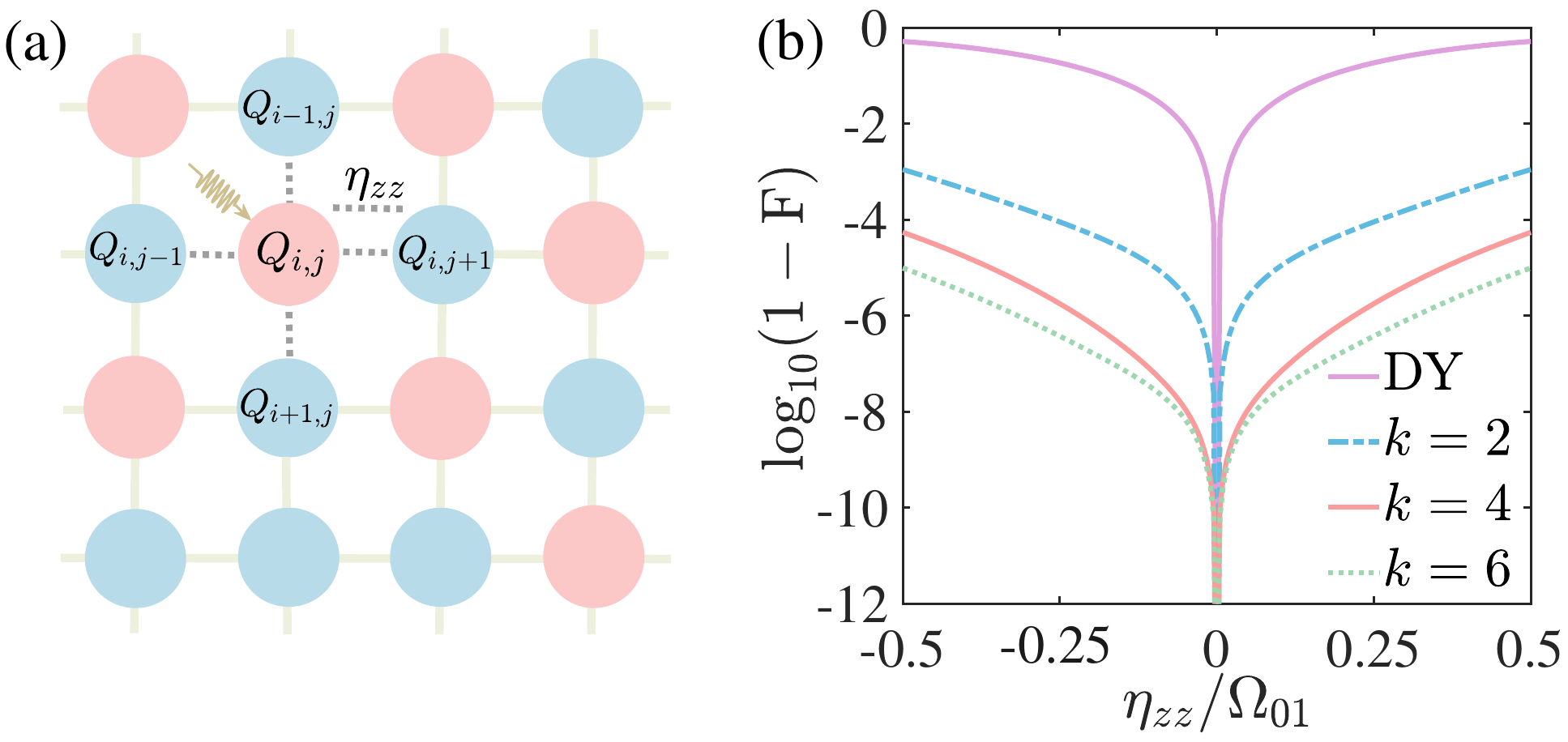}
  \caption{Construction of single-qubit gate with ZZCM. (a) Schematic illustration on a two-dimensional $N\times N$ physical qubits lattice. (b) Gate infidelity of  $\sigma^x/2$ gate on the qubit $Q_{i,j}$ as a function of $\eta_{zz}/\Omega_{01}$ for different values of $k$. The solid purple line represents the result obtained from the dynamical (DY) scheme. }
  \label{lattice}
\end{figure}

We consider a general two-dimensional lattice model for scalable quantum computation. Here, scalable quantum computation refers to the ability to efficiently apply our scheme to  large-scale quantum systems, requiring only local control, in the presence of residual connections. For demonstration purposes and without loss of generality, we assume the lattice consists of $N\times N$ physical qubits, as depicted in Fig. \ref{lattice}(a). In this lattice, we label the qubit at the $i$th row and $j$th column as $Q_{i,j}$. For typical solid-state quantum systems, two nearest-neighbor qubits are coupled through the $XY$ type of interaction \cite{Imamog1999,Jurcevic2014}, and each qubit can be driven independently. The static $ZZ$ coupling is considered as one of the dominant sources of noise \cite{Buterakos2018, Throckmorton2022, AKandala2021, PZhao2022}. The dynamics of the system is governed by the total Hamiltonian $H_{t}(t)=H_{0}(t)+H_{zz}$, with $H_{0}(t) =H_{d}(t)+H_{J}(t)$, where $H_{d}(t)$ represents the Hamiltonian with individual single-qubit drives, $H_{J}(t)$ describes the interaction between nearest-neighbor qubits of the system, and $H_{zz}$ accounts for the $ZZ$-crosstalk Hamiltonian. Setting $\hbar$ = 1 hereafter, in the interaction picture, the Hamiltonian of a single-qubit under resonant driven is given by
\begin{eqnarray}
\label{all Hd}
H_d(t)=\sum_{i, j}\Omega_{i,j} (t)(\cos\phi_{i,j} \sigma_{i,j} ^x+\sin\phi_{i,j} \sigma_{i,j} ^y),
\end{eqnarray}
where $\Omega_{i,j} (t)$ ($\phi_{i,j} $) are the time-dependent driving amplitude (phases) acting on the $Q_{i,j}$ individually, and $\bm{\sigma_{i,j} }=(\sigma_{i,j} ^x, \ \sigma_{i,j} ^y, \ \sigma_{i,j} ^z)$ is Pauli operator. The $XY$ interaction  between nearest-neighbor qubits  is 
\begin{eqnarray}\label{all Hj}
H_{J}(t)&=&\sum_{i, j} \frac{J_{i,j}^x(t)}{2} (\sigma_{i,j}^x \sigma_{i+1,j}^x +\sigma_{i,j}^y \sigma_{i+1,j}^y)\notag \\
&&+\sum_{i, j} \frac{J_{i,j}^y(t)}{2} (\sigma_{i,j}^x \sigma_{i,j+1}^x +\sigma_{i,j}^y \sigma_{i,j+1}^y),
\end{eqnarray}
where $\{i, j\} \in \{1, 2, ...N\}$  and $J_{i,j}^x(t)$ and $J_{i,j}^y(t)$ being the controllable coupling strength between  $Q_{i,j}$ and  its nearest-neighbor qubits along the row and column direction, respectively. The adjacent $ZZ$-crosstalk Hamiltonian is
\begin{eqnarray}
\label{all Hz}
H_{zz}=\sum_{i, j} \left(\eta_{i,j}^x\sigma_{i,j}^z \sigma_{i+1,j}^z + \eta_{i,j}^y\sigma_{i,j}^z \sigma_{i,j+1}^z\right),
\end{eqnarray}
where $\eta_{i,j}^{x,y}$ characterize the coupling strength of $ZZ$ interactions between nearby qubits.

\subsection{The general scheme}

To eliminate the unwanted $ZZ$ crosstalk,
we rotate the system to the framework defined by a time-dependent unitary transformation $\mathcal{A}(t)$ as
\begin{eqnarray}
\label{Hr}
H_{\mathcal{A}}(t)&=&\mathcal{A}^{\dagger}(t)H_{t}(t)\mathcal{A}(t)+{\rm i}\Dot{\mathcal{A}}^{\dagger}(t)\mathcal{A}(t).
\end{eqnarray}
Our goal is to devise the form of $\mathcal{A}(t)$ such that the resulting evolution operator of $H_{\mathcal{A}}(t)$ yields an ideal gate operation at the final moment, i.e.,
\begin{eqnarray}
U_{\mathcal{A}}(T,0) =\mathcal{T}e^{-{\rm i}\int^{T}_{0}H_{\mathcal{A}}(t)dt}=U_{0}
\end{eqnarray}
where $\mathcal{T}$ denotes time ordering, $T$ represents the duration of the gate operation, and $U_{0}$ signifies the ideal gate operation free from the influence of $ZZ$ crosstalk. By imposing the boundary condition of $\mathcal{A}(T)=\mathcal{A}(0)=\bm{I}$, the evolution operator generated by $H_{t}(t)$ in the interaction picture can be expressed as
\begin{eqnarray}
\label{Ut}
U_{t}(T,0)&=&\mathcal{A}(T)U_{\mathcal{A}}(T,0)=\bm{I}\cdot U_{0},
\end{eqnarray}
which implies that we eliminate the adverse effect of $ZZ$ crosstalk and realize the ideal gate operation.


To determine the specific form of $\mathcal{A}(t)$ satisfying Eq. (\ref{Ut}), we divide the evolution into $k$ segments ($k$ is a positive integer), that is $T=k \tau$ with $k$ being positive integer, and expand $U_{\mathcal{A}}(T,0)$ as
\begin{eqnarray}
\label{UR}
U_{\mathcal{A}}(T,0)&=&\prod_{n=1}^{k}U_{\mathcal{A}}[n \tau, (n-1)\tau].
\end{eqnarray}
For the $n$th period $U_{\mathcal{A}}[n\tau, (n-1)\tau]=\mathcal{T} {\rm exp} $ $\left [-{\rm i}\int_{(n-1)\tau}^{n\tau}H_{\mathcal{A}}(t)dt\right ]$,
by using the Magnus expansion \cite{Magnus1954,Blanes2009}, the unitary evolution
operator $U_{\mathcal{A}}[n\tau, (n-1)\tau]$ corresponding to a time-dependent Hamiltonian is  
 \begin{eqnarray}
\label{u0r1}
U_{\mathcal{A}}[n\tau, (n-1)\tau]= e^{ -{\rm i}\tau(\Bar{H}^{(0)}+\Bar{H}^{(1)}+...)  },
\end{eqnarray}
where $\Bar{H}^{(j)}$ is characterized by an order of $\tau^j$. Here, we restrict   to the first-order term $\Bar{H}^{(0)}$, which is known as a lowest-order approximation. It is noteworthy that this approximation becomes exact in the limit of $\tau$ is infinitesimal. In practical situations, we impose the condition that $\tau$ is significantly smaller than the gate time. In the lowest-order approximation, the evolution operator becomes $U_{\mathcal{A}}[n\tau, (n-1)\tau]= {\rm exp}^{ -{\rm i}\tau\Bar{H}^{(0)}  }$, with
 \begin{eqnarray}
\label{Hhat}
&&\Bar{H}^{(0)}=\frac{1}{\tau}\int_{(n-1)\tau}^{n\tau}H_{\mathcal{A}}(t)dt  \\
&&=\frac{1}{\tau}\left [\int_{(n-1)\tau}^{n\tau}H_{\mathcal{A}0}(t)dt+\int_{(n-1)\tau}^{n\tau}\mathcal{A}^{\dagger}(t)H_{zz}\mathcal{A}(t)dt \right ],\notag
\end{eqnarray}
where
 \begin{eqnarray}
\label{Hgg0}
H_{\mathcal{A}0}(t)=\mathcal{A}^{\dagger}(t)H_{0}(t)\mathcal{A}(t)+{\rm i}\Dot{\mathcal{A}}^{\dagger}(t)\mathcal{A}(t)
\end{eqnarray}
is the target  Hamiltonian  without $ZZ$ crosstalk for $H_{0}(t)$, in the $\mathcal{A}(t)$ framework, which can realize the ideal gate  $U_0$.

To eliminate the influence of $H_{zz}$, we impose two conditions to $\mathcal{A}(t)$, i.e.,
\begin{subequations}
\label{Rt}
\begin{align}
&\mathcal{A}(t)=\mathcal{A}(t+\tau), \\
&\int_{0}^{\tau}\mathcal{A}^{\dagger}(t)H_{zz}\mathcal{A}(t)dt=0,
\end{align}
\end{subequations}
which indicate that $\mathcal{A}(t)$ is periodic with its period being $\tau$, and the integral of $H_{zz}$ in the framework of $\mathcal{A}(t)$ is zero, within any periods.
By these settings, Eq. (\ref{UR}) becomes
\begin{eqnarray}
\label{ured}
U_{\mathcal{A}}(T,0) \approx \prod_{n=1}^{k}e^{-{\rm i}\int_{(n-1)\tau}^{n\tau}H_{\mathcal{A}0}(t)dt}  \approx  U_{0},
\end{eqnarray}
which is the ideal rotation operation. Imposing  the boundary condition $\mathcal{A}(T)=\mathcal{A}(0)=\bm{I}$, and  move  back to the interaction picture, we obtain the evolution operator generated by $H_{t}(t)$ is $U_{t}(T,0)=\mathcal{A}(T)U_{\mathcal{A}}(T,0)=U_{0}$. Overall, the key to mitigating $ZZ$ crosstalk and realizing ideal quantum gates is to find a transformed operator $\mathcal{A}(t)$ that satisfies the boundary condition $\mathcal{A}(T)=\mathcal{A}(0)=\bm{I}$ and Eq. (\ref{Rt}).

\section{Examples of universal quantum gates}

In this section, we  exemplify the construction of  universal quantum gates with ZZCM in the considered lattice model, which can support scalable universal quantum computation.  While our simulations are based on a simplified qubit case,  techniques for mitigating  gate error  induced by low anharmonicity of qubits can also be incorporated in our scheme.

\subsection{Single-qubit gate}

As shown in Fig. \ref{lattice}(a), we utilize the construction of a single-qubit $\sigma^x/2$ gate on qubit $Q_{i,j}$ as an example to provide a detailed explanation of how to eliminate $ZZ$ crosstalk from the surrounding four spectator qubits, i.e., $Q_{i-1,j}$, $Q_{i+1,j}$, $Q_{i,j-1}$, $Q_{i,j+1}$. We start from Eq. (\ref{Hgg0}), where the single-qubit driven Hamiltonian in the framework of $\mathcal{A}_1(t)$ reads
\begin{eqnarray}
\label{Hg}
H_{\mathcal{A}1}(t)=\Omega_{01} \left(t \right)\sigma_{i,j}^x,
\end{eqnarray}
where $\Omega_{01} \left(t \right)=\Omega_{01}\sin^2\left(\pi t/T_1\right)$ with $\Omega_{01}$ being the pulse amplitude, and $T_1$ being the gate operation time satisfied $\int_0^{T_1}\Omega_{01}(t)dt=\pi/4$.  Note   that, our scheme does not impose limitations on the shape of $\Omega_{01}(t)$. Here, we chose $\Omega_{01}(t)$ to be a commonly used time-dependent waveform in experimental settings. This waveform exhibits a continuous transition from zero to zero, minimizing abrupt changes, and has practical advantages in experimental implementation, such as reducing undesirable transients.
The nearest $ZZ$-crosstalk Hamiltonian can be written as
\begin{eqnarray}
\label{Hzz1}
H_{zz1}=\eta_{zz}\sigma_{i,j}^z Z_{i, j},
\end{eqnarray}
where
$$Z_{i, j}=\sigma_{i-1,j}^z+\sigma_{i+1,j}^z+\sigma_{i,j-1}^z+\sigma_{i,j+1}^z$$
are the summation of the $\sigma^z$ operators of the four nearest-neighbor qubits for $Q_{i, j}$, and  we assume that the $ZZ$-crosstalk strengths are the same for the sake of simplicity.

For elimination of $ZZ$ crosstalk, we choose the time-dependent transformed operator $\mathcal{A}_1(t)$ to be
\begin{eqnarray}
\label{RA}
\mathcal{A}_1(t)=\exp\left[-{\rm i}\frac{\omega_1 \tau_1}{\pi}\sin^2 \left( \frac{\pi t}{\tau_1}\right) \sigma_{i,j}^{x} \right],
\end{eqnarray}
where $\tau_1=T_1/k$ is the period, and $\omega_1$ being the parameter used to satisfy Eq. (\ref{Rt}b). It is obvious that $\mathcal{A}_1(t)$ satisfies the boundary condition $\mathcal{A}_1(0)=\mathcal{A}_1(T_1)=\bm{I}$ and Eq. (\ref{Rt}a). 
In addition, the condition in Eq. (\ref{Rt}b) can be written as
\begin{eqnarray}
\label{zzint}
&&\int_{(n-1)\tau_1}^{n\tau_1}\mathcal{A}_{1}^{\dagger}(t)H_{zz1}\mathcal{A}_{1}(t)dt  \notag \\
&&=\eta_{zz}\int_{(n-1)\tau_1}^{n\tau_1}(\cos \chi \sigma_{i,j}^{z}+\sin \chi \sigma_{i,j}^{y})Z_{i, j} dt,
\end{eqnarray}
where $\chi=2\omega_{1}  \tau_{1}/\pi\sin^2( \pi t/\tau_{1})$. We define the error cumulant during the $n$th period as
\begin{eqnarray}
\label{error accumulation}
 {\rm EC}=\eta_{zz}\left[\left|\int_{(n-1)\tau_1}^{n\tau_1}\cos \chi  dt \right| +\left|\int_{(n-1)\tau_1}^{n\tau_1}\sin \chi dt \right|\right].
\end{eqnarray}
Through the numerical simulation, it can be concluded that $\omega_1=4.81 k \Omega_{01}$ is the optimal choice for making the error cumulant to be zero, see Appendix \ref{appendix a} for details. 
Note that, as shown in Eq. (\ref{error accumulation}),
$\eta_{zz}$ appears outside of the integrals. Therefore, as long as the integrals  evaluated to be zero over the duration of interest, the  error cumulant will be zero, i.e., independent on the value of $\eta_{zz}$ for different qubits. The assumption of equal $\eta_{zz}$ for all neighbors is set only for the sake of simplicity and clear representation. Meanwhile, Eq. (\ref{error accumulation}) is applicable for the slowly varying drift case, i.e., $\eta_{zz}\ll 1/\tau_1$, this assumption is held for solid-state quantum systems \cite{ZCNi2022,Bermudez2011}.

Once the form of $\mathcal{A}_1(t)$ is determined, we can obtain the total Hamiltonian $H_{1}(t)$ in the interaction picture by inverting Eq. (\ref{Hr}), i.e.,
\begin{eqnarray}
\label{H10}
H_{1}(t) &=&\mathcal{A}_1(t)H_{\mathcal{A}1}(t)\mathcal{A}_{1}^{\dagger}(t)+{\rm i}\frac{d\mathcal{A}_1(t)}{dt}\mathcal{A}_{1}^{\dagger}(t)+H_{zz1}\notag \\
&=&\Omega_1(t)\sigma_{i,j}^x+H_{zz1},
\end{eqnarray}
with  $\Omega_1(t)=\Omega_{01}\sin^2( \pi t/T_1)+\omega_1 \sin(2\pi t/\tau_1)$. Note that, to meet the conditions in Eq. (\ref{Rt}) and the boundary condition $\mathcal{A}(T)=\mathcal{A}(0)=\bm{I}$, the form of is $\mathcal{A}_1(t)$ is not fixed, we choose the sine square form is aimed to set the added waveform, the second term in $\Omega_1(t)$, to be  simple sine function.
This indicates that we can mitigate the $ZZ$ crosstalk from all spectators only by modulating the shape of the external  drive on the gate qubit from $\Omega_{01}(t)$ to $\Omega_1(t)$, and the term ${\rm i} \Dot{\mathcal{A}}_1(t)\mathcal{A}_{1}^{\dagger}(t)$ serves as the additional correction Hamiltonian, which consists of only single qubit terms.

It is worth noting that the target ZZCM Hamiltonian in Eq. (\ref{Hgg0}) is consistent with the type of quantum gate under consideration.
However, on constructing other types of quantum gates, the target ZZCM Hamiltonian in Eq. (\ref{Hgg0}) will be changed accordingly. In this case, the  form of Eq. (\ref{RA}) for $\mathcal{A}_1(t)$ will also be changed in order to suppress  $ZZ$ crosstalk. After setting the target ZZCM Hamiltonian and it corresponding $\mathcal{A}_1(t)$,  the resulting Hamiltonian in the interaction picture can still be obtained using the formula in the first line of Eq. (\ref{H10}). Generally, the design of $\mathcal{A}(t)$ needs careful consideration to ensure that the Hamiltonian in the interaction picture, namely Eq. (\ref{H10}), possesses concise form and retains practical feasibility. 

 We numerically quantify the gate robustness by using the gate fidelity of $F(U_0)=|Tr(U_0^{\dagger}U_{zz})|/|Tr(U_0^{\dagger}U_0)|$ \cite{XGWang2009}, where $U_0$ and $U_{zz}$ are the ideal and error-affected evolution operators, respectively.
Figure \ref{lattice}(b) shows a comparison of the robustness to $ZZ$ crosstalk of ZZCM schemes with different $k$ and the DY scheme, where the DY scheme is implemented by using a pulse of $\Omega_{d1}(t)=\Omega_{01}\sin^2{(\pi t/T_{d1})}$, with $T_{d1}$ being the gate time, see Appendix \ref{appendix f} for details. The numerical results reveal a positive correlation between an increase in $k$ and an enhancement of the resilience to $ZZ$ crosstalk. This relationship stems from the fact that, larger $k$ corresponds to smaller $\tau_1 = T_1/k$, which leads to an improved precision of the lowest-order Magnus expansion approximation. This heightened precision in representing the system's dynamics results in a more efficient mitigation of $ZZ$ crosstalk. Additionally, we can observe that the gate infidelity as a function of $\eta_{zz}/\Omega_{01} $ ($\eta_{zz}/\Omega_{01} \in [-0.5,0.5]$) can be smaller than $10^{-4}$ throughout the entire range of errors for $k\geq4$. Notably, compared to the DY scheme, the gate infidelity of the ZZCM schemes reduces by at least 2 orders of magnitude when the error ratio exceeds $|0.02|$.

  It is noted that,  from a general theoretical perspective, it is true that smaller values of $\tau_1$ (or larger $k$) lead to a more accurate lowest-order approximation of Magnus expansion, which, in turn, results in a more effective suppression of $ZZ$ crosstalk. However, we can infer from Eq. (\ref{H10}) that the modulated coupling strength $\Omega_1(t)$ increases as $k$ increases, which is not favorable for implementation in an actual experimental setup.  
Therefore, when selecting $k$, one needs to balance the $ZZ$ crosstalk suppression and  experimentally feasibility for the control pulses. This balance will depend on the specific physical system under investigation. To address this issue, we set the maximum value of $\Omega_1(t)$ as $\Omega_m$, with $\eta_{zz}/\Omega_m$ ranging from $-0.05$ to 0.05. Under these conditions, the optimal choice of $k$ is found to be 4, as shown in Appendix \ref{appendix b}.


 \begin{figure}[tbp]
  \centering
  \includegraphics[width=\linewidth]{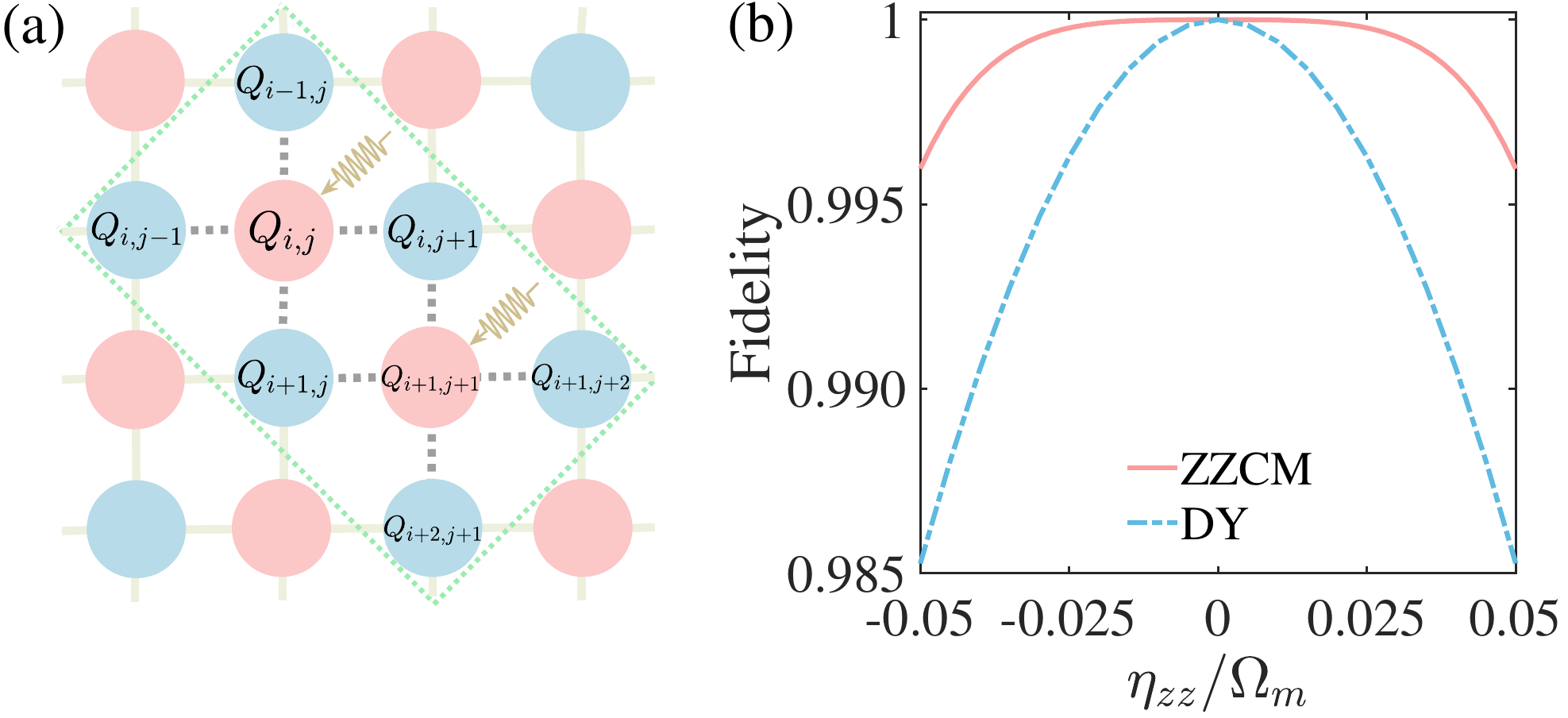}
  \caption{Construction of parallel single-qubit gate with ZZCM. (a) The lattice structure for constructing  single-qubit gates $\sigma^x_{i,j}\otimes\sigma^y_{i+1,j+1}$, i.e., $\sigma^x$ on qubit $Q_{i,j}$ and $\sigma^y$ on qubit $Q_{i+1,j+1}$. (b) Gate fidelity of parallel gates as a function of $\eta_{zz}/\Omega_m$.
  }
  \label{singlequbit}
\end{figure}


\subsection{Parallel single-qubit gates}
In the context of scalable quantum computation, parallel quantum gates play a pivotal role. 
In the  absence of $ZZ$ crosstalk, simultaneously constructed single-qubit gates are trivial, which can be induced by independent driving fields. However, in the presence of $ZZ$ crosstalk, parallel single-qubit gates are no longer trivial, as manipulating one of the qubits will also lead to unintended operation on its nearby qubits, leading them  to be  entangled with gate qubits. Therefore, we now focus on the suppression of interqubit $ZZ$ crosstalk during the construction of parallel single-qubit gates. 

Here we take the parallel single-qubit gates $\sigma^x$ on $Q_{i,j}$ and $\sigma^y$ on $Q_{i+1,j+1}$ simultaneously as an example, as shown in the dotted box in Fig. \ref{singlequbit}(a). Note that, parallel single-qubit gates on nearest-neighbor qubits can also be implemented, as detailed in Appendix \ref{appendix c}.
Starting from Eq. (\ref{Hgg0}), the individual single-qubit driven Hamiltonian is
\begin{eqnarray}
H_{\mathcal{A}2}(t)=\Omega_{02}(t)(\sigma_{i,j}^x+\sigma_{i+1,j+1}^y)
\end{eqnarray}
with $\Omega_{02}(t)=\Omega_{02}\sin^2( \pi t/T_2)$, where the gate operation time $T_{2}$ satisfies $\int_0^{T_{2}}\Omega_{02}(t) dt=\pi/2$.
In this case, the Hamiltonian of nearest $ZZ$ crosstalk can be written as
\begin{eqnarray}\label{hzz2}
H_{zz2}=\eta_{zz}(\sigma_{i,j}^z Z_{i,j}+\sigma_{i+1,j+1}^z Z_{i+1,j+1}).
\end{eqnarray}
When we choose the transformed operator as
 \begin{eqnarray}
\label{A2}
\mathcal{A}_2(t)=  {\rm exp}\left [-{\rm i}\frac{\omega_{2}  \tau_{2}}{\pi} \sin^2 \left( \frac{\pi t}{\tau_{2}}\right)(\sigma_{i,j}^{x}+\sigma_{i+1,j+1}^{y})\right ],
\end{eqnarray}
 the Hamiltonian in the interaction picture can be obtained by inverting Eq. (\ref{Hr}), i.e.,
\begin{eqnarray}
  H_2(t)= \Omega_2(t)(\sigma_{i,j}^{x}+\sigma_{i+1,j+1}^{y} )+H_{zz2},
 \end{eqnarray}
where the equivalent coupling strength is $\Omega_2(t)=\Omega_{02}\sin^2( \pi t/T_{2})+\omega_{2}\sin(2\pi t/\tau_{2})$. Here, $\tau_{2}=T_{2}/k$ is the period, and $\omega_2$ is the parameter used to satisfy Eq. (\ref{Rt}b).
By utilizing numerical simulations, we can determine the optimal value of $\omega_2$ is $2.4 k\Omega_{02}$, which ensures the  error cumulant to be zero, see Appendix \ref{appendix a} for details. By setting the maximum value of $\Omega_2(t)$ to $\Omega_m$, and imposing a constraint on $\eta_{zz}/\Omega_m$ in the range of $[-0.05, 0.05]$, we can identify the optimal value of $k$ to be 4.
A comparison between the robustness of the ZZCM scheme with $k=4$ and the DY scheme against $ZZ$ crosstalk is presented in Fig. \ref{singlequbit}(b), where the DY scheme is implemented by using a pulse of $\Omega_{d2}(t)=\Omega_{m}\sin^2({\pi t/T_{d2}})$, with $T_{d2}$ being the gate time, see Appendix \ref{appendix f} for details. The results indicate a meaningful increase in robustness for the entire error range when implementing the ZZCM proposal.



 \begin{figure}[tbp]
  \centering
  \includegraphics[width=\linewidth]{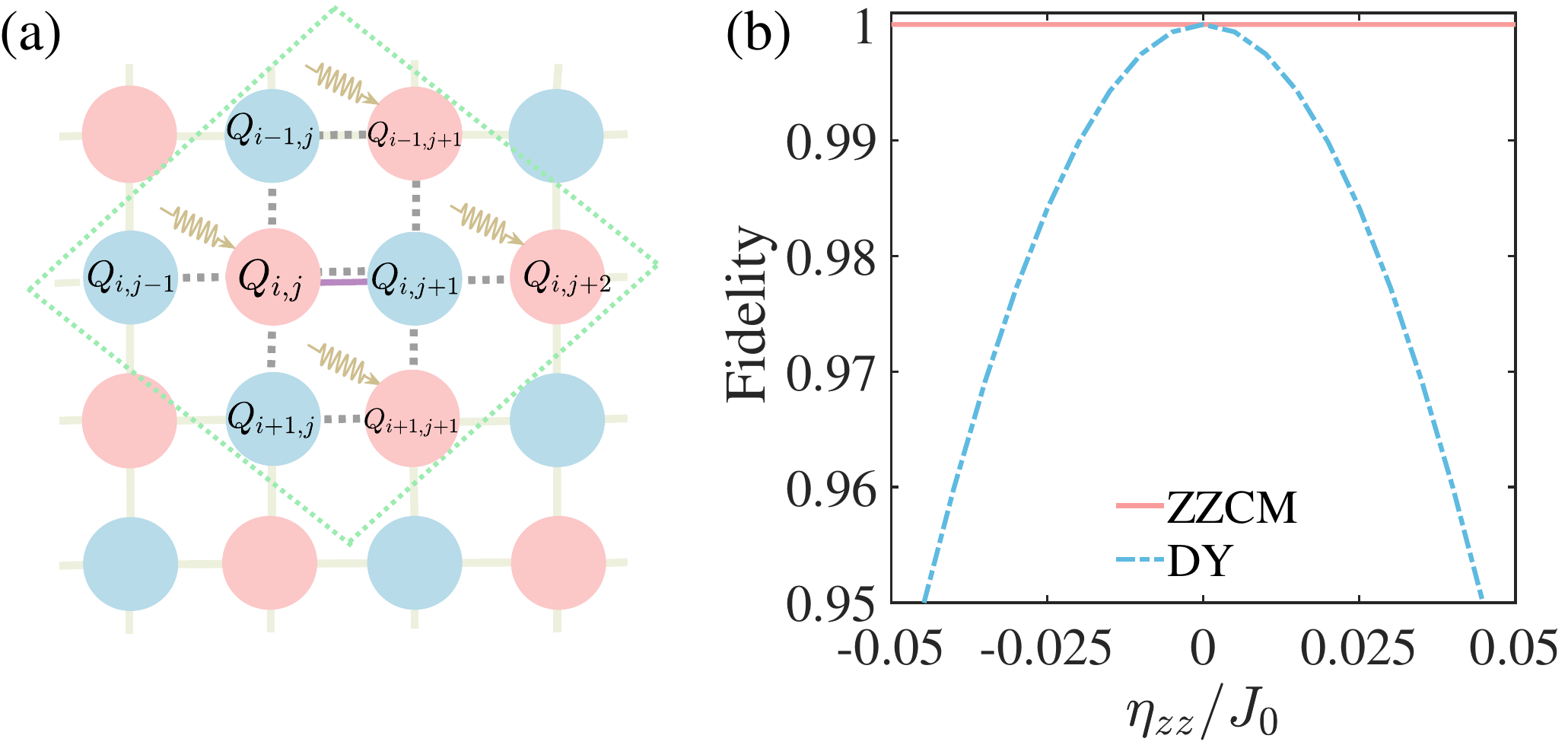}
  \caption{Simultaneously construction of  single- and two-qubit gates with ZZCM.  (a) The lattice structure for constructing  gates of $U_{{(i,j)},{(i,j+1)}}^{S}\otimes \sigma^x_{{i-1,j+1}}\otimes \sigma^y_{{i,j+2}}\otimes I_{{i+1,j+1}}$, which construct a SWAP gate on qubits $Q_{i,j}$ and $Q_{i,j+1}$, a $\sigma^x$ gate on qubit $Q_{i-1,j+1}$, a $\sigma^y$ gate on qubit $Q_{i,j+2}$, and the identity operation on qubit $Q_{i+1,j+1}$, simultaneously. (b) Gate fidelity  as a function of $\eta_{zz}/J_0$.  
  }
  \label{zzduibi}
\end{figure}

\subsection{The two-qubit SWAP gate}

Compared to single-qubit gates, two-qubit gates are more susceptible to parasitic $ZZ$ crosstalk, significantly challenging the attainment of high-performance two-qubit gates. Therefore, suppressing $ZZ$ crosstalk is crucial for implementing high-performance two-qubit gates. In this regard, we devise the two-qubit SWAP gate with $ZZ$-crosstalk mitigation effects. Considering a quantum system consisting of eight physical qubits enclosed in a dotted box in Fig. \ref{zzduibi}(a), where the SWAP gate $U_{(i,j),(i,j+1)}^S$ are acted on qubits $Q_{i,j}$ and $Q_{i,j+1}$, and the remaining six qubits, $Q_{i-1,j},Q_{i-1,j+1},Q_{i,j-1},Q_{i,j+2},Q_{i+1,j},Q_{i+1,j+1}$, denote the spectators. The interaction between  qubits is indicated by the solid and dashed lines, which correspond to the $XY$ and $ZZ$ interactions, respectively. 

It is worth noting that, unlike the approach taken for constructing single-qubit gates, when building two-qubit gates, we follow the general principles outlined in Sec. \ref{sec2}, i.e., we start with the Hamiltonian in the interaction picture. When constructing single-qubit quantum gates, we start from the target Hamiltonian in Eq. (\ref{Hg}), in the framework of $\mathcal{A}_1(t)$,  to obtain the  interaction  Hamiltonian in Eq. (\ref{H10}) in a reverse way. This reverse approach allows us to establish the relationship between the Hamiltonian in the interaction picture  and the rotation operator $\mathcal{A}(t)$. This  approach is applicable  for single-qubit gate construction, as arbitrary direction of single-qubit control  is experimental achievable. 
However, when constructing two-qubit gates, the form of the two-qubit coupling  is in specific forms for different quantum systems. Here, as shown in Fig. \ref{zzduibi}(a), we consider the conventional $XY$ interactions between nearby qubits, and use only single-qubit operations to suppress $ZZ$ crosstalk. Therefore, we initially set the form of  the interaction Hamiltonian,  to avoid introducing two-qubit interactions beyond the intended $XY$ coupling.  Inspired by Eq. (\ref{H10}), we represent the interaction  Hamiltonian as 
 \begin{eqnarray}
\label{H3}
H_3(t)=H_J(t)+H_{zz3}+H_{A3}(t),
\end{eqnarray}
where
\begin{eqnarray}
H_J(t)={J(t) \over 2} (\sigma_{i,j}^{x}\sigma_{i,j+1}^{x}+\sigma_{i,j}^{y}\sigma_{i,j+1}^{y})
\end{eqnarray}
is the $XY$ interaction Hamiltonian between qubits $Q_{i,j}$ and $Q_{i,j+1}$, with $J(t)=J_{0}\sin^2(\pi t/T_{3})$ being the time-dependent interaction strength. The Hamiltonian
\begin{eqnarray}
H_{zz3}=\eta_{zz}[\sigma_{i,j}^z Z_{i,j}+\sigma_{i,j+1}^z (Z_{i,j+1}-\sigma_{i,j}^z)]
\end{eqnarray}
represents the $ZZ$ crosstalk between qubits. The additional Hamiltonian $H_{A3}(t)={\rm i}\dot{\mathcal{A}}_3(t)\mathcal{A}^{\dagger}_3(t)$ is introduced to suppress the $ZZ$ crosstalk.

To achieve the SWAP gate while suppressing $ZZ$ crosstalk by only single-qubit gate correction, we divide the evolution into two steps. In the first step, we choose $\mathcal{A}_{31}(t)= {\rm exp}$ $ \left [-{\rm i} \omega_{3}  \tau_{3}/\pi\sin^2(  \pi t/\tau_{3}) (\sigma_{i-1,j+1}^x+\sigma_{i,j}^x+\sigma_{i,j+2}^x+\sigma_{i+1,j+1}^x)\right ]$, with $\tau_{3}=T_{3}/4$ (meaning $k=4$), and $\omega_{3}= 4\times2.4 J_{0}$. Similar to the singe-qubit case, the choice of $\mathcal{A}_{31}(t)$ is still not fixed, it needs to only satisfy the boundary conditions and Eq. (\ref{Rt}). Here, for the sake of simplicity, we set $\mathcal{A}_{31}$ in a single-qubit rotation form, making the additional correction term, which is $H_{A3}(t)$ in Eq. (\ref{H3}), just contain single-qubit operations.
By rotating the system to the framework defined by $\mathcal{A}_{31}(t)$, we obtain
 \begin{eqnarray}
\label{h3A}
H^1_{\mathcal{A}3}(t)&=&\mathcal{A}_{31}^{\dagger}(t)H_{3}(t)\mathcal{A}_{31}(t)+{\rm i}\Dot{\mathcal{A}}^{\dagger}_{31}(t)\mathcal{A}_{31}(t) \notag \\
&=& \frac{J(t)}{2}\sigma_{i,j}^{x}\sigma_{i,j+1}^{x}\\
&+& \mathcal{A}_{31}^{\dagger}(t)\left[\frac{J(t)}{2}\sigma_{i,j}^{y}\sigma_{i,j+1}^{y}+H_{zz3}\right]\mathcal{A}_{31}(t),\notag
\end{eqnarray}
where $\int_{(n-1)\tau_3}^{n\tau_3}\mathcal{A}_{31}^{\dagger}(t) [ J(t)\sigma_{i,j}^{y}\sigma_{i,j+1}^{y}/2$ $+H_{zz3} ]\mathcal{A}_{31}(t)dt=0$. Hence, when $\int_0^{T_{3}}J(t)dt=\pi/2$, in the framework of $\mathcal{A}_{31}(t)$, the evolution operator in the first step is $U^1_{\mathcal{A}3}(T_3,0)={\rm exp}[-{\rm i} \pi \sigma_{i,j}^{x}\sigma_{i,j+1}^{x}/4 ]$, which represents a nontrivial two-qubit gate, i.e., we can perform two-qubit gate with suppression of $ZZ$ crosstalk in only one step. By combining arbitrary single-qubit gate operations, we can achieve universal quantum computation. However, to enable direct comparison with the dynamical gate scheme, we target  to construct a SWAP gate. This is because under $XY$ interactions, the dynamical gate scheme can directly implement a SWAP gate instead of  $U^1_{\mathcal{A}3}(T_3,0)$. To build the SWAP gate, we need to proceed with the next step.


In the second step, we choose $\mathcal{A}_{32}(t)= {\rm exp}$ $\left [-{\rm i} \omega_{3}  \tau_{3}/\pi\sin^2(  \pi t/\tau_{3})  (\sigma_{i-1,j+1}^y+\sigma_{i,j}^y+\sigma_{i,j+2}^y+\sigma_{i+1,j+1}^y)\right ]$, and the Hamiltonian $H_{A3}(t)$ in Eq. (\ref{H3}), in the framework defined by $\mathcal{A}_{32}(t)$, is
 \begin{eqnarray}
\label{h3A2}
H^2_{\mathcal{A}3}(t)&=&\mathcal{A}_{32}^{\dagger}(t)H_{3}(t)\mathcal{A}_{32}(t)+{\rm i}\Dot{\mathcal{A}}^{\dagger}_{32}(t)\mathcal{A}_{32}(t) \notag \\
&=&\frac{J(t)}{2}\sigma_{i,j}^{y}\sigma_{i,j+1}^{y}\\
&+& \mathcal{A}_{32}^{\dagger}(t)\left[\frac{J(t)}{2}\sigma_{i,j}^{x}\sigma_{i,j+1}^{x}+H_{zz3}\right]\mathcal{A}_{32}(t).\notag
\end{eqnarray}
Since $\int_{(n-1)\tau_3}^{n\tau_3}\mathcal{A}_{32}^{\dagger}(t)\left [ J(t)\sigma_{i,j}^{x}\sigma_{i,j+1}^{x}/2+H_{zz3}\right]\mathcal{A}_{32}(t)dt=0$, the evolution operator in the second step becomes $U^2_{\mathcal{A}3}(2T_3,T_3)={\rm exp}[-{\rm i} \pi \sigma_{i,j}^{y}\sigma_{i,j+1}^{y}/4 ]$. As a result, the evolution operator of the whole process is
\begin{eqnarray}
U_{\mathcal{A}3}(2T_3,0)=U_{\mathcal{A}3}(2T_3,T_3)U_{\mathcal{A}3}(T_3,0)=U^S_{(i,j),(i,j+1)},\notag \\
\end{eqnarray}
which is a SWAP gate between qubits $Q_{i,j}$ and $Q_{i,j+1}$. Moving back to the interaction picture, the evolution operator generated by $H_3(t)$ at the end time reads
 \begin{eqnarray}
U_3(2T_3)=\mathcal{A}_{32}(2T_3)U^S_{(i,j),(i,j+1)}=U^S_{(i,j),(i,j+1)}.
 \end{eqnarray}
Therefore, we realize the two-qubit SWAP gate with ZZCM.

It should be noted that, to suppress $ZZ$ crosstalk, an additional Hamiltonian $H_{A3}(t)={\rm i}\dot{\mathcal{A}}_3(t)\mathcal{A}^{\dagger}_3(t)$ is introduced, as shown in Eq. (\ref{H3}). This extra Hamiltonian requires additional drives not only on qubit $Q_{i,j}$ but also on spectator qubits $Q_{i-1,j+1}$, $Q_{i,j+2}$, and $Q_{i+1,j+1}$. We emphasize that these spectator qubits do not represent any additional physical qubits overhead on the quantum processor, as we can still perform independent quantum gate operations on them. 
This means that while implementing  the SWAP gate, parallel  single-qubit gates on the three qubit can also be obtained simultaneously.
To demonstrate this, we introduce a parallel gate $U_{{(i,j)},{(i,j+1)}}^{S}\otimes \sigma^x_{{i-1,j+1}}\otimes \sigma^y_{{i,j+2}}\otimes I_{{i+1,j+1}}$, which consists of a SWAP gate on qubits $Q_{i,j}$ and $Q_{i,j+1}$, a  $\sigma^x$ gate on qubit $Q_{i-1,j+1}$, a $\sigma^y$ gate on qubit $Q_{i,j+2}$, and an identity gate on qubit $Q_{i+1,j+1}$. In this scenario, the total Hamiltonian is represented as $H'_3(t)=H_3(t)+H_s(t)$, where
\begin{eqnarray}
H_s(t)={J(t) \over 2} (\sigma_{i- 1,j+1}^{x}+\sigma_{i,j+2}^{y})
 \end{eqnarray}
is the Hamiltonian for constructing the single-qubit gates.

Figure \ref{zzduibi}(b) displays the numerically simulated gate fidelity as a function of $ZZ$ crosstalk. The comparison between the performance of ZZCM and DY schemes is presented, where the latter is implemented via the Hamiltonian of
\begin{eqnarray}
\label{d3}
H_{d3}(t)&=&{J(t) \over 2}(\sigma_{i,j}^{x}\sigma_{i,j+1}^{x}+\sigma_{i,j}^{y}\sigma_{i,j+1}^{y}) \notag\\
&&+ J(t) (\sigma_{i-1,j+1}^{x}+\sigma_{i,j+2}^{y}),
\end{eqnarray}
see Appendix \ref{appendix f} for details. Figure \ref{zzduibi}(b) demonstrates that the ZZCM scheme exhibits remarkable robustness against $ZZ$ crosstalk, outperforming the DY scheme. The results reveal that the two-qubit gate is much more sensitive to $ZZ$ crosstalk than the single-qubit gate and that the incorporation of the ZZCM method can efficiently mitigate this issue. Specifically, the incorporation of the ZZCM approach reduces the infidelity of the parallel gate $U_{{(i,j)},{(i,j+1)}}^{S}\otimes \sigma^x_{{i-1,j+1}}\otimes \sigma^y_{{i,j+2}}\otimes I_{{i+1,j+1}}$ by 3 orders of magnitude, compared to the DY scheme, when the $ZZ$ crosstalk ratio is 0.05.
Parallel SWAP gates can also be implemented, see Appendix \ref{appendix e} for details.


\section{DISCUSSION AND CONCLUSION } We have presented a protocol  for implementing $ZZ$-crosstalk-mitigation  universal quantum gates in a  two-dimensional square lattice. This approach is also applicable to other kinds of lattices, encompassing three-dimensional qubit arrays. It eliminates the need for auxiliary qubits, simplifying the implementation process and reducing resource requirements. Moreover, the method is compatible with different types of quantum processors, accommodating both direct and bus-based qubit interactions. These features contribute to the versatility and scalability of the $ZZ$-crosstalk mitigation scheme, making it a promising approach for a wide range of quantum computation platforms.
By employing a time-dependent unitary transformation operator, we successfully realize high-performance isolated and parallel quantum gates while mitigating the $ZZ$ crosstalk between qubits. Notably, the ZZCM proposal can be utilized to prevent the accumulation and propagation of errors induced by $ZZ$ crosstalk, making it a promising solution for constructing deep quantum circuits and simulating quantum algorithms. Consequently, our protocol may lay the groundwork for practical, scalable, and fault-tolerant quantum computation.

\section*{acknowledgements}
This work was supported by the National Natural Science Foundation of China (Grant No. 12275090 and No. 12304554), the Guangdong Provincial Key Laboratory (Grant No. 2020B1212060066), the Innovation Program for Quantum Science and Technology (Grant No. 2021ZD0302303), NSFC/RGC JRS (Grant No. N-HKU774/21), and the CRF of Hong Kong (Grant No. C6009-20G).

\appendix

 \section{Numerical Simulation for the Error cumulant} \label{appendix a}

\begin{figure}[t]
  \centering
  \includegraphics[width=\linewidth]{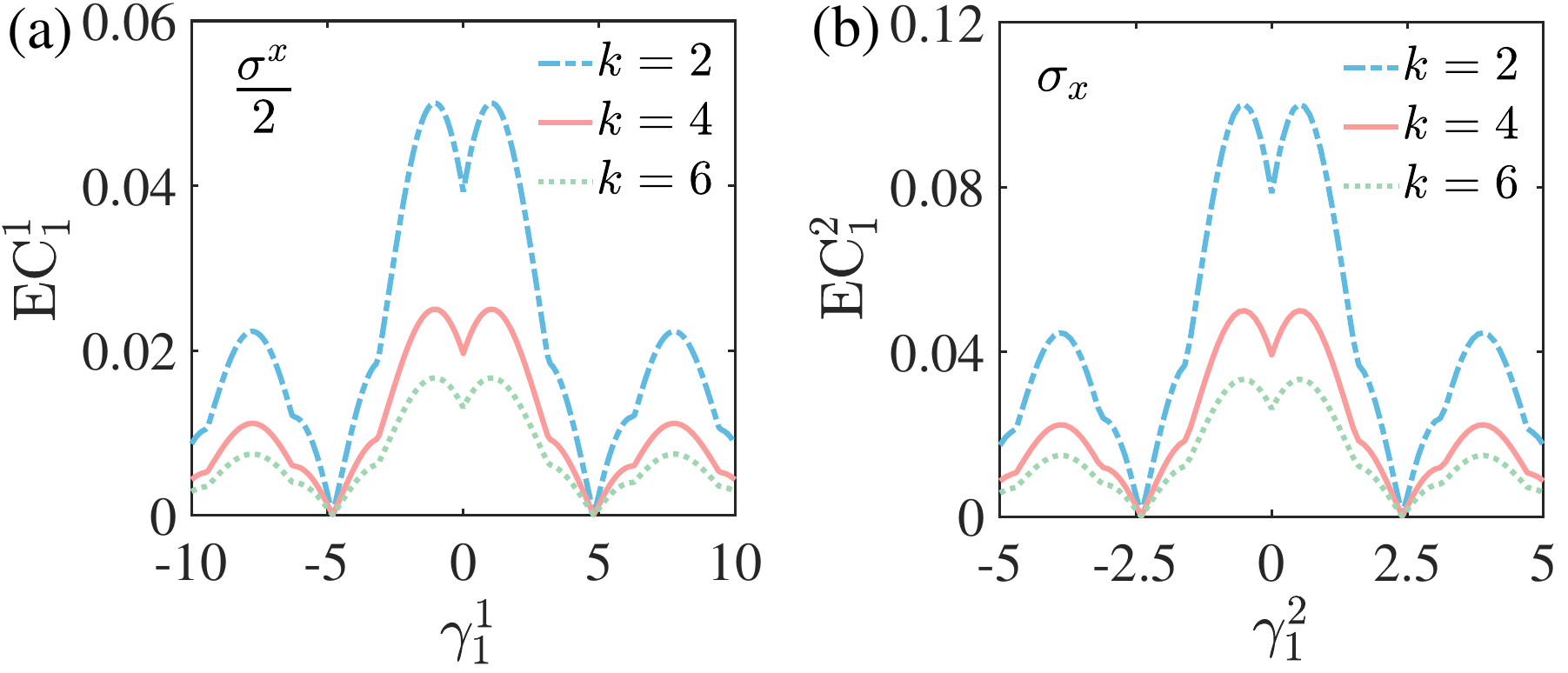}
  \caption{The error cumulant as a function of $\gamma_1^l$, where the optimal values of $\gamma_1^l$ for isolated $\sigma^x/2$ and $\sigma^x$  gates are $\gamma_1^1=|4.81|$ and $\gamma_1^2=|2.4|$, respectively,  when  setting $\eta_{zz}=0.05\Omega_{01}$.  }
  \label{figs1}
\end{figure}

To mitigate the influence of $ZZ$ crosstalk, the time-dependent transformed operator $\mathcal{A}(t)$ needs to satisfy the requirement stated in Eq. (\ref{Rt}) of the main text.
For single-qubit gates $U_{(l=1,2)}$ ($U_1=\sigma^x/2, \ U_2=\sigma^x$) on qubit $Q_{i,j}$, the corresponding operation duration satisfy $\int_0^{T_{1}^1}\Omega_{01}\sin^2( \pi t/T_{1}^1)dt=\pi/4$ and $\int_0^{T_1^{2}}  \Omega_{01}\sin^2( \pi t/T_1^{2})dt=\pi/2$, respectively. The Hamiltonian describing the $ZZ$ crosstalk between nearest qubits reads in Eq. (\ref{Hzz1}).
To mitigate the $ZZ$ crosstalk from nearest qubits, we choose $\mathcal{A}_1^l(t)={\rm exp}\left [-{\rm i} \omega_1^l \tau_1^l/\pi \sin^2( \pi t/\tau_1^l) \sigma_{i,j}^{x} \right ] $, where $\tau_1^l=T_1^l/k$ is the period, and $\omega_1^{l}=\gamma_1^l k \Omega_{01} $ ($k$ is positive integer) is the parameter used to satisfy Eq. (\ref{Rt}b). Therefore, the condition in Eq. (\ref{Rt}b) can be expressed as
\begin{eqnarray}
\label{Azzint1}
&&\int_{(n-1)\tau_{1}^{l}}^{n\tau_1^{l}}\mathcal{A}_{1}^{l\dagger}(t)H_{zz1}\mathcal{A}_1^{l}(t)dt \notag \\
&&=\eta_{zz} \int_{(n-1)\tau_1^l}^{n\tau_1^l}(\cos \chi_1^l \sigma_{i,j}^{z}+\sin \chi_1^l \sigma_{i,j}^{y}) Z_{i, j} dt,
\end{eqnarray}
where $\chi_1^l=2\gamma_1^l k \Omega_{01}\tau_1^{l}/\pi\sin^2( \pi t/\tau_1^{l})$. We define the error cumulant during the $n$th period as
\begin{eqnarray}
\label{AEC1}
{\rm EC}_1^l=\eta_{zz}\left[\left|\int_{(n-1)\tau_1^l}^{n\tau_1^l}\cos\chi_1^l dt \right| +\left|\int_{(n-1)\tau_1^l}^{n\tau_1^l}\sin\chi_1^l dt \right|\right]. \notag \\
\end{eqnarray}
Figures~\ref{figs1}(a) and (b) display the error cumulant as a function of $\gamma_1^l$, indicating that for the $\sigma^x/2 \ (\sigma^x)$ gate, $\gamma_1^1=|4.81| \ (\gamma_1^2=|2.4|)$ is the optimal choice for eliminating the $ZZ$ crosstalk.

\begin{figure}[t]
  \centering
  \includegraphics[width=\linewidth]{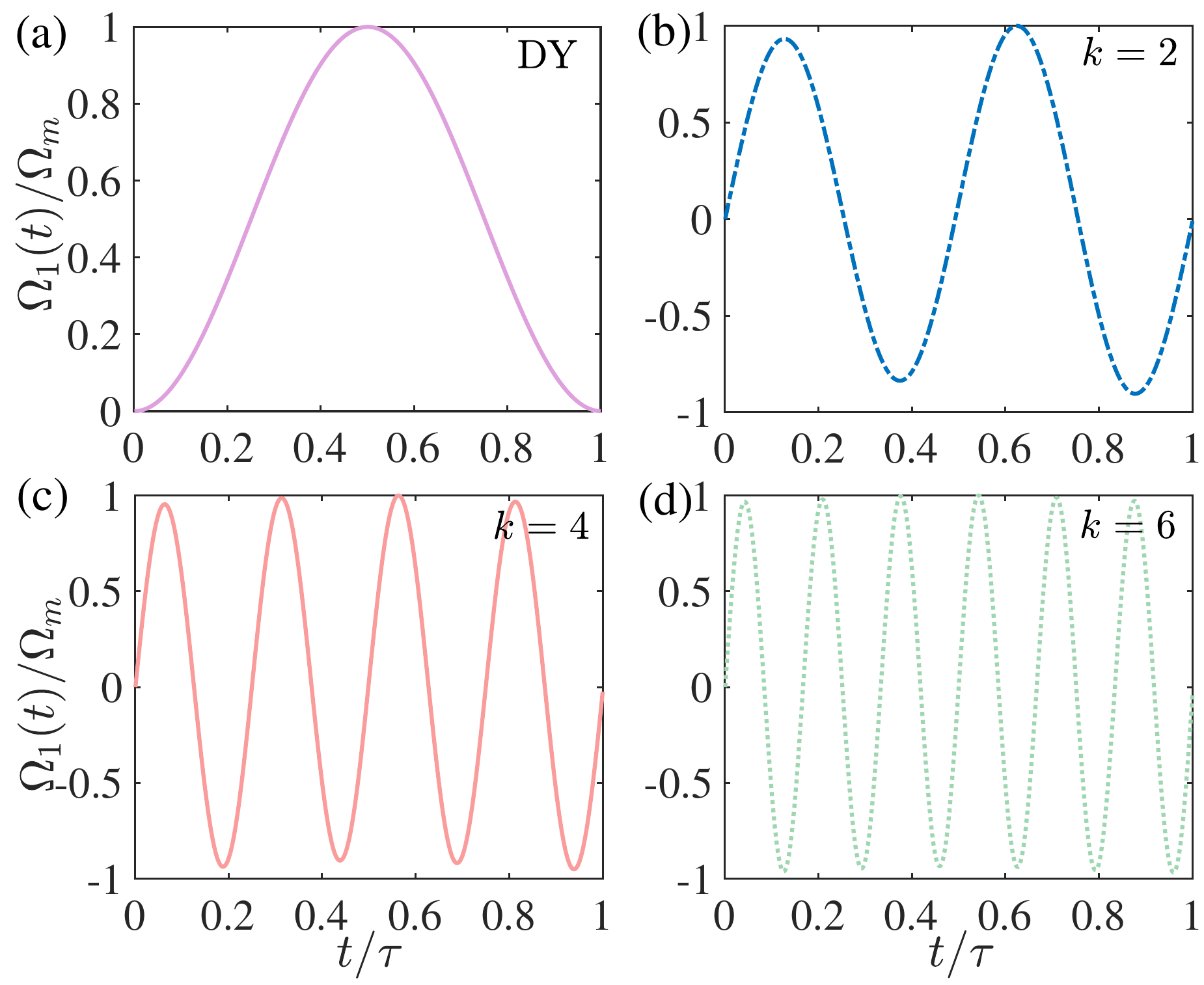}
  \caption{ The waveform of the modulated  drive $\Omega_1(t)$ for the $\sigma^x/2$ gate, (a) in  the dynamical gate scheme and (b)-(d) in our ZZCM scheme with $\gamma_1^1=4.81$ and different $k$. }  
  \label{fig_rabi}
\end{figure}

After setting $\gamma_1$,   the shape of the modulated amplitude  $\Omega_1(t)$ for different values of $k$ can be determined, as shown in Fig. \ref{fig_rabi}. The modulated  $\Omega_1(t)$ is presented in Eq. (\ref{H10}), with $\Omega_1(t)=\Omega_{01}\sin^2( \pi t/T_1)+\omega_1 \sin(2\pi t/\tau_1)$, and $\omega_1=\gamma_1 k \Omega_{01} $. As $\omega_1$ is proportional to  $k$, the effective maximum amplitude of $\Omega_1(t)$ will be larger than the original driving field. Despite this limitation, as numerically verified, the gate performance can still be greatly improved. In addition, the modulated drive  maintains the waveform of the original field as a simple periodic function and can be readily obtained experimentally  by  arbitrary waveform generators \cite{Cundiff2010,Takase2022}.

For the parallel single-qubit gate on next-nearest-neighbor qubits $Q_{i,j}$ and $Q_{i+1,j+1}$, the $ZZ$-crosstalk Hamiltonian is in the form in Eq. (\ref{hzz2}).
The gate operation time $T_{2}$ satisfies $\int_0^{T_{2}}\Omega_{02}\sin^2( \pi t/T_{2})dt=\pi/2$. When we choose the transformed operator in the form in Eq. (\ref{A2}), with $\tau_{2}=T_{2}/k$ being the period, and $\omega_{2}=\gamma_{2} k \Omega_{02}$ used to satisfy Eq. (\ref{Rt}b). In this case, the condition in Eq. (\ref{Rt}b) can be written as
\begin{eqnarray}
\label{Azzint2}
&&\int_{(n-1)\tau_{2}}^{n\tau_{2}}\mathcal{A}_{2}^{\dagger}(t)H_{zz2}\mathcal{A}_{2}(t)dt \notag \\
&&=\eta_{zz} [\int_{(n-1)\tau_{2}}^{n\tau_{2}}(\cos \chi_{2} \sigma_{i,j}^{z}+\sin \chi_{2} \sigma_{i,j}^{y}) Z_{i,j}dt \\
&&
+\int_{(n-1)\tau_{2}}^{n\tau_{2}}(\cos \chi_{2} \sigma_{{i+1,j+1}}^{z} -\sin \chi_{2} \sigma_{i+1,j+1}^{x})Z_{i+1,j+1} dt ],\notag
\end{eqnarray}
with $\chi_{2}=2\gamma_{2} k \Omega_{02}\tau_{2}/\pi\sin^2( \pi t/\tau_{2})$. We can also define the error cumulant during the $n$th period as
\begin{eqnarray}
\label{AEC2}
{\rm EC_2}=\eta_{zz}\left[\left|\int_{(n-1)\tau_2}^{n\tau_2}\cos\chi_2 dt \right| +\left|\int_{(n-1)\tau_2}^{n\tau_2}\sin\chi_2 dt \right|\right], \notag \\
\end{eqnarray}
which is consistent with the error cumulant in isolated $\sigma^x$ gate.

\section{The construction of dynamical gates} \label{appendix f}

Here, we present the  construction of dynamical (DY) gates with simple resonant interaction. They are implemented by using time-dependent driving fields, and the corresponding Hamiltonians are  $H_{d1}(t)= \Omega_{d}\sin^2( \pi t/T_{d1}) \sigma_{i,j}^x $ and $H_{d2}(t)=\Omega_{d}\sin^2( \pi t/T_{d2})(\sigma_{i,j}^x+\sigma_{i',j'}^y)$ for single-qubit gate on qubit $Q_{i,j}$, and parallel single-qubit gate $\sigma^x_{i,j}\otimes \sigma^y _{i',j'}$ between nearest or next-nearest neighboring qubits $Q_{i,j}$ and $Q_{i',j'}$, respectively. To construct the isolated gate $\sigma^x/2$ ($\sigma^x$), the corresponding evolution time $T_{d1}$ satisfies $\int_0^{T_{d1}}\Omega_{d}\sin^2( \pi t/T_{d1})dt=\pi/4 \ (\pi/2)$. Similarly, for the parallel single-qubit gate, the evolution time is   $\int_0^{T_{d2}}\Omega_{d}\sin^2( \pi t/T_{d2})dt= \pi/2 $. The amplitude of the driven field is set to be equal to the amplitude of the equivalent coupling strength in the ZZCM scheme, i.e., $ \Omega_d=\Omega_m$.

We also present the implementation of the two-qubit SWAP gate using a simple $XY$ interaction. The corresponding Hamiltonian is given by
\begin{eqnarray}
H_{d3}(t)=J_d(t) (\sigma_{i,j}^{x}\sigma_{i',j'}^{x}+\sigma_{i,j}^{y}\sigma_{i',j'}^{y})/2,
\end{eqnarray}
where $J_d(t)=J_{0}\sin^2(\pi t/T_{d3})$ is the time-dependent $XY$ interaction between qubits $Q_{i,j}$ and $Q_{i',j'}$, and the evolution time satisfies $\int_0^{T_{d3}}J_d(t)dt=\pi/2$.

Moreover, we construct parallel gates of $U_{(i,j),(i,j+1)}^{S}\otimes \sigma^x_{i-1,j+1}\otimes \sigma^y_{i,j+2}\otimes I_{i+1,j+1}$ and $U_{(i,j),(i,j+1)}^{S}\otimes U_{(i+1,j),(i+1,j+1)}^{S}$ by using the DY method. The corresponding Hamiltonian is given by Eq. (\ref{d3})
and
\begin{eqnarray}
H_{d5}(t) &= & {J_d(t) \over 2} (\sigma_{i,j}^{x}\sigma_{i,j+1}^{x}+\sigma_{i,j}^{y}\sigma_{i,j+1}^{y}\notag\\
&&+\sigma_{i+1,j}^{x}\sigma_{i+1,j+1}^{x}+\sigma_{i+1,j}^{y}\sigma_{i+1,j+1}^{y}),
\end{eqnarray}
respectively, where the evolution time is chosen to satisfy $\int_0^{T_{d4(5)}}J_d(t)dt=\pi/2$. We note that the form of the $ZZ$-crosstalk Hamiltonian is the same as in the ZZCM scheme.

 \begin{figure}[t]
  \centering
  \includegraphics[width= \linewidth]{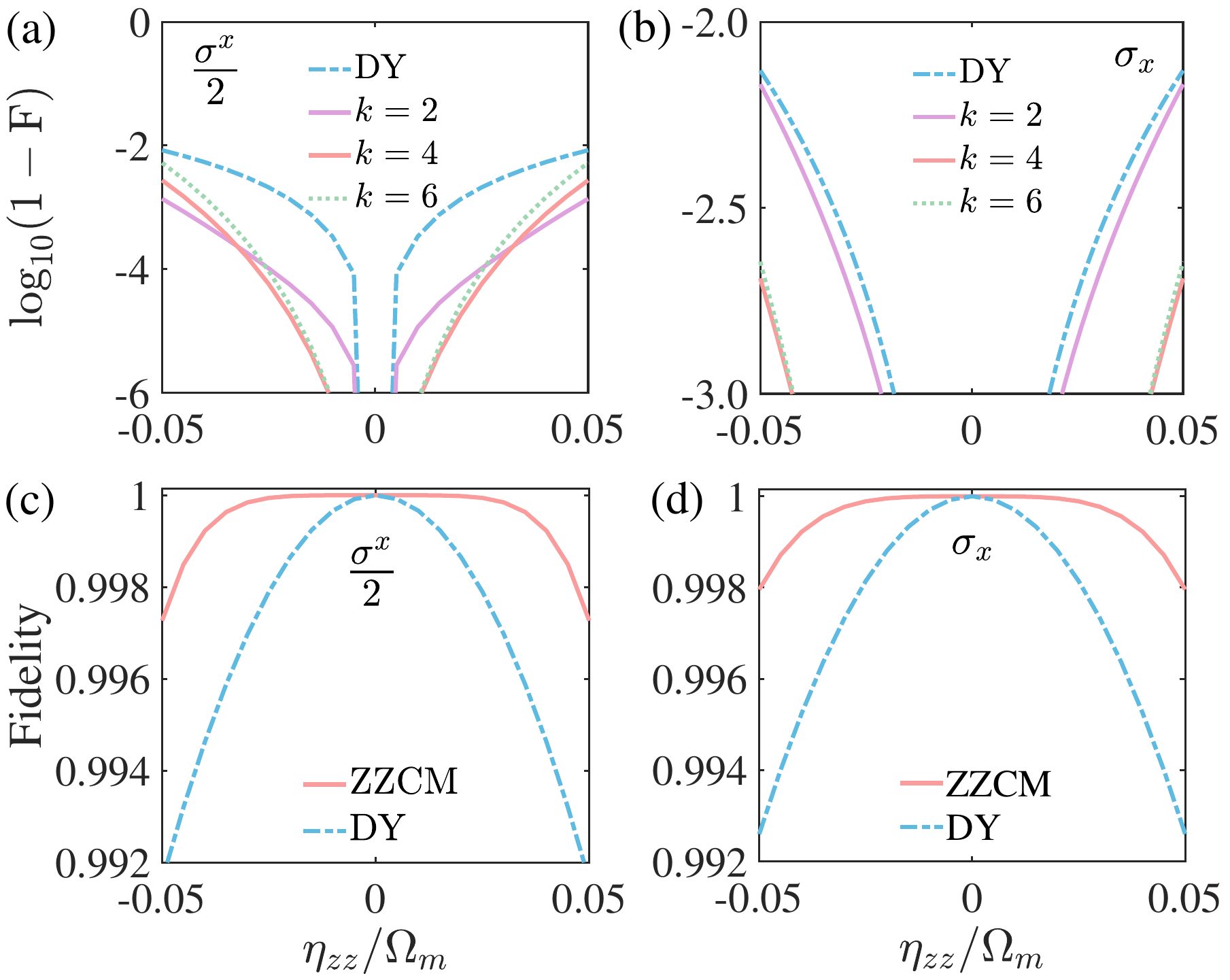}
  \caption{ (a) and (b) display the infidelity of isolated $\sigma^x/2$ and $\sigma^x$ gates for various $k$ under the condition that the maximum equivalent coupling strength is $\Omega_m$, as a function of $\eta_{zz}/\Omega_m$. (c) and (d) compare the robustness of isolated $\sigma^x/2$ and $\sigma^x$ gates of the ZZCM scheme with $k=4$ and the DY scheme. }
  \label{figs2}
\end{figure}

\section{The Optimal $k$ for the ZZCM Scheme} \label{appendix b}

We draw attention to the incorporation of the ZZCM control, which induces an increase in the effective coupling strength as $k$ grows, as established by Eq. (\ref{H10}) in the main paper. The corresponding equivalent coupling strength for the ZZCM approach is provided by $\Omega_1(t)=\Omega_{01}\sin^2( \pi t/T_{1})+\omega_{1}\sin(2\pi t/\tau_{1})$, where $\omega_{1}=\gamma_1 k \Omega_{01}$. To prevent excessive driving field amplitude in experiments, the amplitude of the equivalent coupling strength is constrained to $\Omega_m$, and the $ZZ$ crosstalk ratio $\eta_{zz}/\Omega_m$ is within the range of $[-0.05,0.05]$. Under these conditions, as $k$ increases, the gate-driven field amplitude $\Omega_{01}$ diminishes, consequently resulting in increased relative noise, quantified by the parameter $\eta_{zz}/\Omega_{01}$. Therefore, there exists a trade-off between the enhancement in robustness with increasing $k$ and the proportion of noise $\eta_{zz}/\Omega_{01}$.  Figures \ref{figs2}(a) and (b) depict the infidelities of $\sigma^x/2$ and $\sigma^x$ gates, respectively, as a function of $\eta_{zz}/\Omega_m$, with an optimal value of $k=4$ identified for the ZZCM scheme. Additionally, Figs. \ref{figs2}(c) and (d) contrast the robustness of the ZZCM scheme, with $k=4$, against the DY scheme concerning $ZZ$ crosstalk when the equivalent coupling strength amplitude is $\Omega_m$. The plots distinctly showcase the outstanding suppression of $ZZ$ crosstalk achieved by the ZZCM scheme.

 \begin{figure}[t]
  \centering
  \includegraphics[width= \linewidth]{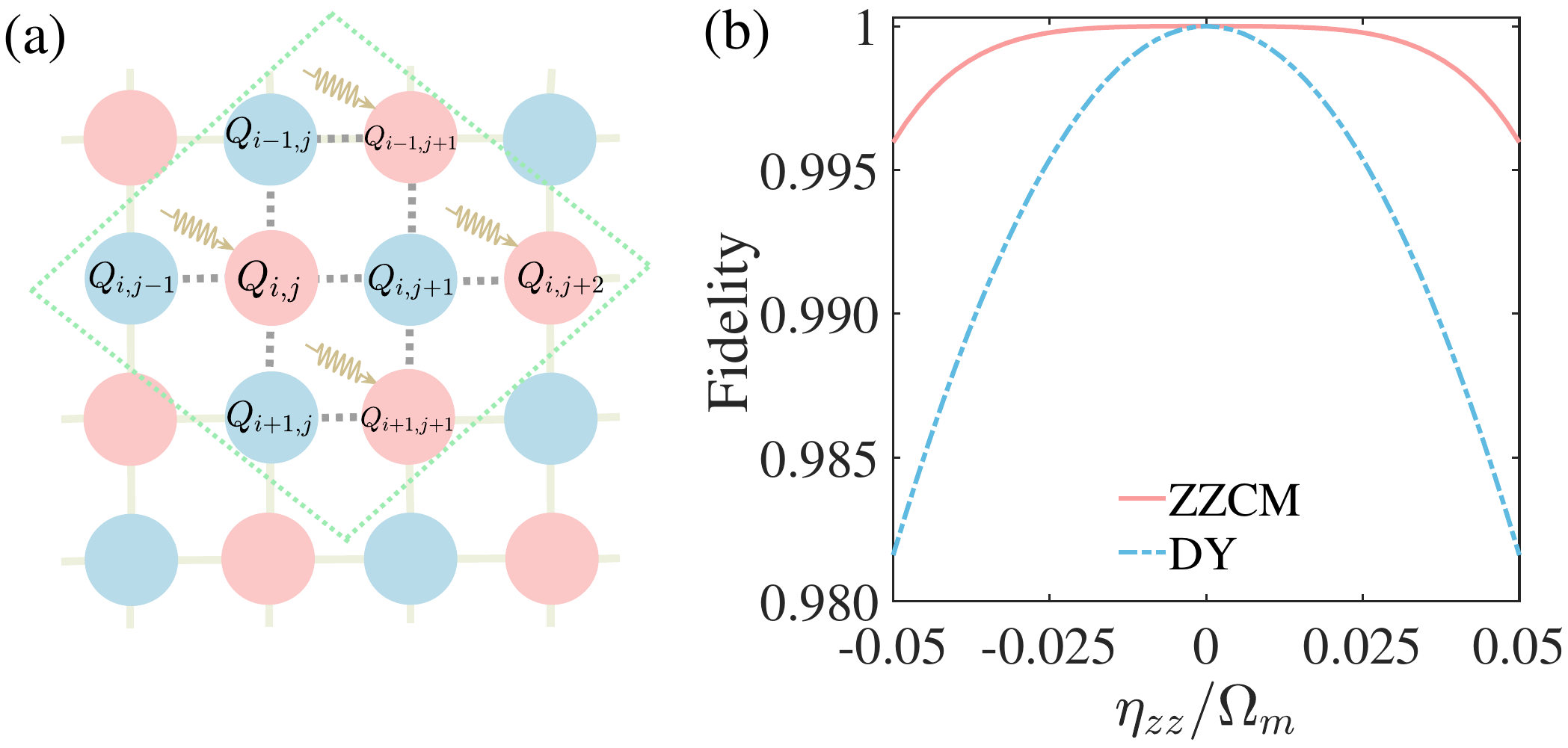}
  \caption{Implementation of  parallel single-qubit gates. (a) Schematic diagram for  $\sigma^x_{i,j} \otimes \sigma^y_{i,j+1}$ gates between nearest qubits $Q_{i,j}$ and $Q_{i,j+1}$. (b) compares the robustness of the parallel gate $\sigma^x_{i,j} \otimes \sigma^y_{i,j+1}$ between the ZZCM scheme with $k=4$ and the DY scheme. }
  \label{figs3}
\end{figure}

\section{The construction of parallel gates on nearby qubits} \label{appendix c}

Here, we present the construction of parallel single-qubit gate between nearest-neighbor qubits in an eight-qubit system, which is enclosed by the dotted box in Fig. \ref{figs3}(a). We apply $\sigma^x$ and $\sigma^y$ gates simultaneously on qubits $Q_{i,j}$ and $Q_{i,j+1}$, and the remaining six qubits serve as spectators. Starting from Eq. (\ref{Hgg0}) in the main text, the individual single-qubit-driven Hamiltonian in the $\mathcal{A}$ framework is $H_{\mathcal{A}2}'(t)=\Omega_{02}\sin^2( \pi t/T_2)(\sigma_{i,j}^x+\sigma_{i,j+1}^y)$, and the $ZZ$ interaction Hamiltonian is
\begin{eqnarray}
H_{zz2}'=\eta_{zz}[\sigma_{i,j}^z Z_{i,j}+\sigma_{i,j+1}^z(Z_{i,j+1} -\sigma_{i,j}^z)].
\end{eqnarray}
To suppress the $ZZ$ crosstalk between qubits, we choose
\begin{eqnarray}
\mathcal{A}_2'(t) &=&  {\rm exp}
\left [-{\rm i} {\omega_{2}  \tau_{2} \over \pi}
\sin^2 \left ({ \pi t\over \tau_{2}}\right) (\sigma_{i-1,j+1}^{x} \right.\notag\\
&&+\left.\sigma_{i,j}^{x}+\sigma_{i,j+2}^{x}+\sigma_{i+1,j+1}^{x})\right ].
\end{eqnarray}
The total Hamiltonian,
\begin{eqnarray}
H_2'(t)&=&   H_{zz2} + \Omega_2(t)\sigma_{i,j}^{x}
+\Omega_{02}\sin^2 \left ({ \pi t\over T_{2}}\right)  \sigma_{i,j+1}^y\notag\\
&& +\Omega_{\mathcal{A}}(\sigma_{i-1,j+1}^{x}+\sigma_{i,j+2}^{x}+\sigma_{i+1,j+1}^{x}),
\end{eqnarray}
in the interaction picture can be obtained by inverting Eq. (4) of the main text, where $\Omega_{\mathcal{A}}=\omega_{2}\sin(2\pi t/\tau_{2})$, and $\Omega_2(t)=\Omega_{02}\sin^2( \pi t/T_{2})+\Omega_{\mathcal{A}}$ is the equivalent coupling strength applying on qubit $Q_{i,j}$. It indicates that, for mitigating $ZZ$ crosstalk, we need additional control fields apply on qubits $Q_{i-1,j+1}$, $Q_{i,j}$, $Q_{i,j+2}$, and $Q_{i+1,j+1}$. However, it is worth noting that qubits $Q_{i-1,j+1}$, $Q_{i,j+2}$, and $Q_{i+1,j+1}$ do not incur extra resource consumption on the quantum processor as we can still implement single-qubit gates independently on these qubits.

 We conduct a comparative analysis of the ZZCM scheme and the DY scheme, focusing on their robustness against $ZZ$ crosstalk. We here impose a constraint on the maximum value of the equivalent coupling strength $\Omega_2(t)$, limiting it to $\Omega_m$. Comparing Fig. \ref{figs3}(b) and Figs. \ref{figs2}(c) and (d), we observe that the susceptibility of the parallel single-qubit gate to $ZZ$ crosstalk exceeds that of isolated single-qubit gates. Fortunately, the proposed ZZCM scheme effectively mitigates the adverse effects of $ZZ$ crosstalk. Consequently, the parallel single-qubit gate achieves comparable performance to isolated single-qubit gates.

\begin{figure}[tb]
	\centering
			\includegraphics[width= \linewidth]{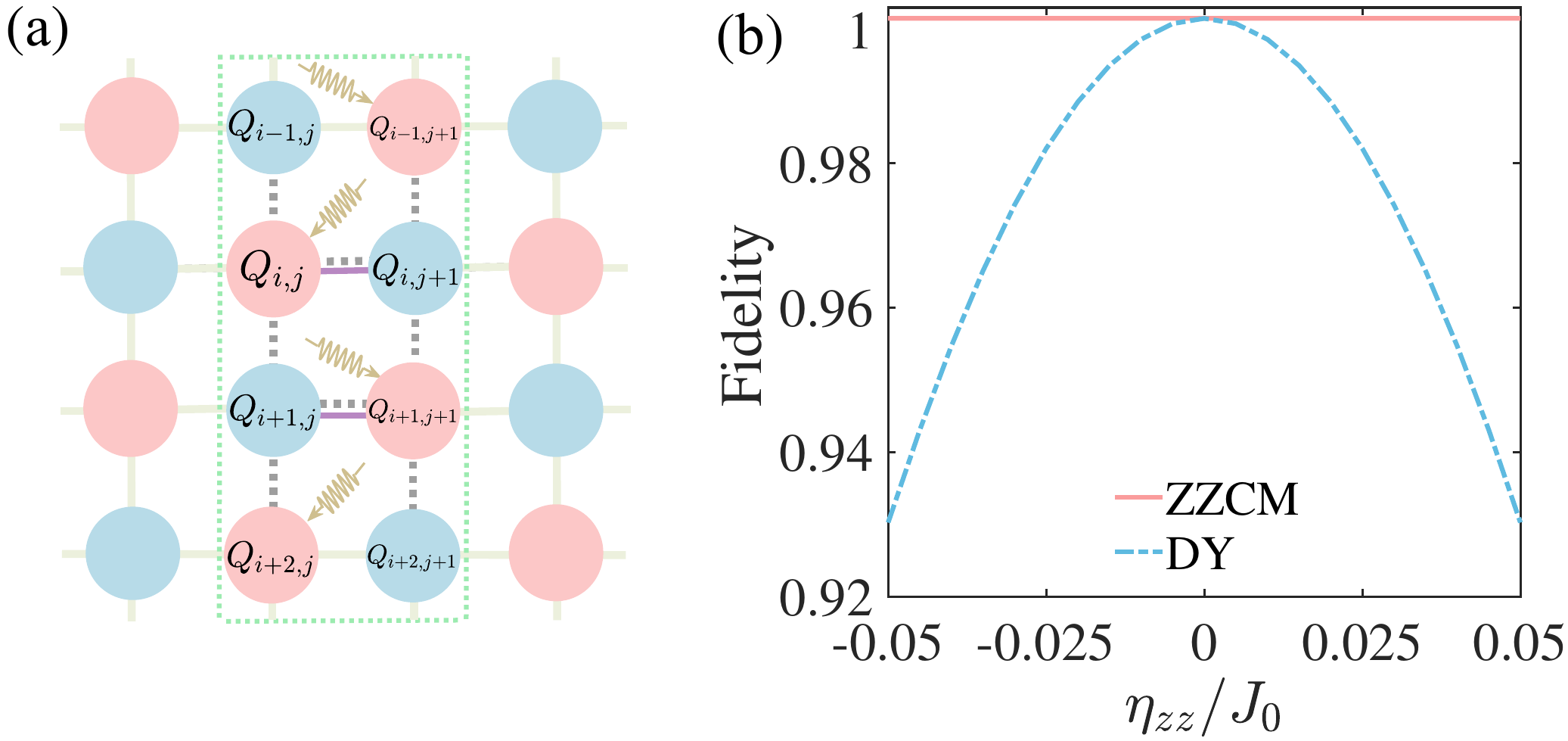}
    	\caption{Implementation of two-qubit gates. (a) shows the schematic diagrams for the parallel gated $U_{(i,j),(i,j+1)}^S\otimes U_{(i+1,j),(i+1,j+1)}^S$. (b) compares the robustness between the ZZCM scheme with $k=4$ and the DY scheme, respectively. }
	\label{figs4}
\end{figure}


\section{The construction of parallel SWAP gates } \label{appendix e}

We next demonstrate the application of the ZZCM control for realizing the parallel two-qubit SWAP gate in an eight-qubit system, as shown in the dotted box in Fig. \ref{figs4}(a).
The SWAP gates on qubits $Q_{i,j}$ and $Q_{i,j+1}$, $Q_{i+1,j}$ and $Q_{i+1,j+1}$ are implemented by turning on the $XY$ interaction between the respective qubit pairs. The total Hamiltonian of this quantum system in the interaction picture is
\begin{eqnarray}
\label{h4}
H_4(t)=H'_J(t)+H_{zz4}+H'_A(t),
\end{eqnarray}
which is composed of the $XY$ interaction Hamiltonian
\begin{eqnarray}
H'_J(t)&=&{J(t)' \over 2} (\sigma_{i,j}^{x}\sigma_{i,j+1}^{x}+\sigma_{i,j}^{y}\sigma_{i,j+1}^{y}\notag\\
&&+\sigma_{i+1,j}^{x}\sigma_{i+1,j+1}^{x}+\sigma_{i+1,j}^{y}\sigma_{i+1,j+1}^{y})
\end{eqnarray}
between qubits $Q_{i,j}$ and $Q_{i,j+1}$, $Q_{i+1,j}$ and $Q_{i+1,j+1}$, with $J(t)'=J_{0}\sin^2(\pi t/T_{4})$ as the time-dependent interaction strength. The $ZZ$-crosstalk Hamiltonian in this system is described by $H_{zz4}=\eta_{zz}$ $[\sigma_{i,j}^z(Z_{i,j}-\sigma_{i,j-1}^z)+\sigma_{i,j+1}^z(\sigma_{i-1,j+1}^z+\sigma_{i+1,j+1}^z)$ $+\sigma_{i+1,j}^z(\sigma_{i+1,j+1}^z+\sigma_{i+2,j}^z)+\sigma_{i+1,j+1}^z\sigma_{i+2,j+1}^z  ]$. To suppress the $ZZ$ crosstalk, we use the additional Hamiltonian $H'_A(t)={\rm i}\dot{\mathcal{A}}_4(t)\mathcal{A}^{\dagger}_4(t)$.

The procedure of constructing the parallel SWAP gate shares similarities with the construction process of an isolated SWAP gate. However, there is a key difference lies in the selection of transformed operator $\mathcal{A}(t)$. Specifically, we select $\mathcal{A}_{41}(t)= {\rm exp}$ $\left [-{\rm i} \omega_{4}  \tau_{4}/\pi\sin^2(  \pi t/\tau_{4}) (\sigma_{i-1,j+1}^{x}+\sigma_{i,j}^{x}+\sigma_{i+1,j+1}^{x}+\sigma_{i+2,j}^{x})\right ]$ in the first step,  followed by $\mathcal{A}_{42}(t)= {\rm exp}$ $\left [-{\rm i} \omega_{4}  \tau_{4}/\pi\sin^2(  \pi t/\tau_{4}) (\sigma_{i-1,j+1}^{y}+\sigma_{i,j}^{y}+\sigma_{i+1,j+1}^{y}+\sigma_{i+2,j}^{y})\right ]$ in the second step, with $\tau_{4}=T_{4}/k$, and $\omega_{4}=2.4 k J_{0}$. Under these settings, the ZZCM scheme exhibits superior robustness against $ZZ$ crosstalk as expected, as shown in Fig. \ref{figs4}(b). In contrast, the gate fidelity experiences a rapid decline as $\eta_{zz}/J_{0}$ increases when the ZZCM control is absent. However, by employing the ZZCM method, we successfully enhance the fidelity of the parallel SWAP gate from an initial value of $93.02\%$ to $99.99\%$ when the $ZZ$-crosstalk ratio is $\eta_{zz}/J_{0}=|0.05|$.

\end{document}